\documentstyle[eqsecnum,preprint,aps,prd,epsf,rotate]{revtex}
\preprint{NSF-ITP-98-077, DAMTP R-1998/77}
\date{\today}
\draft
\tighten
\begin{document}
\def\sqr#1#2{{\vcenter{\hrule height.3pt
      \hbox{\vrule width.3pt height#2pt  \kern#1pt
         \vrule width.3pt}  \hrule height.3pt}}}
\def\square{\mathchoice{\sqr67\,}{\sqr67\,}\sqr{3}{3.5}\sqr{3}{3.5}}
\def\today{\ifcase\month\or
  January\or February\or March\or April\or May\or June\or July\or
  August\or September\or October\or November\or December\fi
  \space\number\day, \number\year}

\def\Bbb{\bf}


\title{Puncture of gravitating domain walls}

\author{Andrew Chamblin{$^{1}$} and Douglas M. Eardley{$^2$}}

\address {\qquad \\ {$^1$}Center for Theoretical Physics, Massachusetts Institute of\\
Technology, Bldg. 6-304, Cambridge, MA 02139, USA
\qquad \\ {$^2$}Institute for Theoretical Physics\\
University of California\\
Santa Barbara, California 93106-4030, U.S.A.
}

\maketitle

\begin{abstract}

We investigate the semi-classical instability of vacuum domain
walls to processes where the domain walls decay by the formation
of closed string loop boundaries on their worldvolumes.  Intuitively, a wall which
is initially spherical may `pop', so that a hole corresponding to a string
boundary component on the wall, may form.  We find instantons,
and calculate the rates, for such processes.  We show that after puncture,
the hole grows exponentially at the same rate that the wall expands.
It follows that the wall is never completely thermalized by a single expanding hole;
at arbitrarily late times there is still a large, thin shell of matter which
may drive an exponential expansion of the universe.  We also study the situation
where the wall is subjected to multiple punctures.  We find that in order
to completely annihilate the wall by this process, at least four string loops
must be nucleated.  We argue that this process may be relevant in certain brane-world
scenarios, where the universe itself is a domain wall.

\end{abstract}


\section{Introduction}

Hybrid topological defects can occur in a variety of physical scenarios 
(\cite{review}, \cite{sean}) where multiple phase transitions are allowed
to occur simultaneously.  Typically, one may envision lower dimensional
defects ending on higher dimensional defects (so-called `Dirichlet'
topological defects \cite{sean}), or conversely higher dimensional defects
may end on lower dimensional defects \cite{kib}.  In this paper, we are
chiefly concerned with the latter situation, where domain walls may have
string boundary components.

In the first papers to study walls ending on strings \cite{kib}, 
the authors considered an 
explicit symmetry breaking pattern whereby the unifying gauge symmetry
${\mbox{Spin}(10)}$ is broken to $SU(3) ~{\times}~ SU(2) ~{\times}~ U(1)$
by way of the group $SU(4) ~{\times}~ SU(2) ~{\times}~ SU(2)$.  The first
symmetry breaking pattern gives rise to the existence of ${\Bbb Z}_{2}$
strings, which may form boundaries for the domain walls which are produced
by the second symmetry breaking.

In fact, this basic picture will hold in any scenario where you have a 
series of phase transitions of the form
\begin{equation}
{{\cal G} ~{\stackrel{X_1}{\longrightarrow}}~ {\cal H}
~{\stackrel{X_2}{\longrightarrow}}~ {I}}
\end{equation}

\noindent where ${\pi}_{0}({\cal G}) = {\pi}_{1}({\cal G}) = I$
and ${\pi}_{0}({\cal H}) ~{\neq}~ I$ (here $I$ denotes the identity group).
Since the sequence (1.1) is exact it follows that 
\[
{\pi}_{1}({\cal G}/{\cal H}) = {\pi}_{0}({\cal H})
\]

\noindent That is to say, the first symmetry breaking give rise to strings.
The second phase transition then gives rise to domain walls.
More explicitly, if we consider the behaviour of the field $X_2$ as
we move around a string, we see that $X_2$ must have a discontinuity
since the field is not invariant under the action of ${\cal H}$.
In other words, we have to `add in' domain walls to account for this
discontinuity.

It follows that domain walls in these multiple phase transition scenarios
are not topologically stable defects.  Instead, such walls are unstable to 
a quantum mechanical decay process, 
whereby a closed string loop boundary component appears on 
the worldvolume of the domain wall, and then expands.  To calculate
the probability, $P$, that this process occurs, one typically invokes the 
semiclassical approximation 
\[
P {\propto} Ae^{-(S_{E} - S_{B})}
\]

\noindent where $S_{E}$ is the Euclidean action of the instanton
of the post-tunneling configuration
(a domain wall with a hole in it), $S_{B}$ is the Euclidean action
of the background instanton (a domain wall without a hole in it), and
$A$ is a prefactor which is calculated by considering fluctuations
about the instanton.  Typically, one ignores the prefactor and focusses
on the action terms (since these terms lead to exponential suppression,
they typically dominate).

The trajectory of a virtual string loop moving in Euclidean space is 
a two-sphere of radius $R$, where $R$ is the radius at which the string
nucleates.  The Euclidean section of a planar domain wall is (topologically)
still just ${\Bbb R}^3$.  Thus, the instanton for the final state is
${\Bbb R}^3$ with a ball of radius $R$ removed, and the instanton for 
the initial state is ${\Bbb R}^3$.  It follows that the difference between
the two actions must contain a boundary term proportional to the area of
the two-sphere, and a bulk term proportional to the volume of the removed
region; indeed, one obtains \cite{review}
\begin{equation}
S_{E} - S_{B} = 4{\pi}R^{2}{\mu} - \frac{4{\pi}}{3}R^{3}{\sigma}
\end{equation}

\noindent where $\mu$ is the string tension and $\sigma$ is the energy density
of the domain wall.  For typical symmetry breaking scales associated with
strings and walls, one finds that $S_{E} - S_{B} >> 1$, and so this decay 
process is usually supressed \cite{review}.

In all of this analysis, we have not said anything about the gravitational
fields of these domain walls.  This is because we have been assuming that 
the walls are so `light' that they have effectively decoupled from gravity.
Of course, there is no reason why heavy domain walls should not suffer
the same instabilities as light walls.  We are therefore led to generalize
the known work on light domain walls, to include the effects of gravity.
However, the gravitational effects of heavy domain walls are highly
non-trivial; indeed, a gravitating domain wall generically closes the universe!
Of course, this overwhelming property of domain walls is just another reason
why we should be interested in finding new decay modes to get rid of them.
We now turn to a short discussion of the gravitational effects of domain walls,
before outlining the new work on how they may decay by the formation of closed
string loop boundary components.

Throughout this paper we use units in which $\hbar = c = G = 1$.

\section{VIS domain walls: A brief introduction}

Solutions for the gravitational field of a domain wall were found
by Vilenkin \cite{vil} (for an open wall) and Ipser and Sikivie \cite{ip}
(for closed walls).  The global structure of these Vilenkin-Ipser-Sikivie
(or `VIS') domain walls has been extensively discussed recently (\cite{cald},
\cite{shawn1}) so we will only present a brief sketch here.

To begin, we look for a solution of the Einstein equations
where the source term is an energy momentum tensor describing a distributional
source located at $z = 0$:
\begin{equation}
T_{\mu \nu} = \sigma \delta(x) {\rm diag} (1,1,1,0) 
\label{stressenergy}
\end{equation}

It is impossible to find a static solution of the Einstein equations with
this source term; indeed, the VIS solution is a time-dependent solution
describing a uniformly accelerating domain wall.  In order to understand
the global causal structure of the VIS domain wall, it is most useful
to use coordinates $(t, x, y, z)$ so that the metric takes the form
\begin{equation} 
ds^{2} = e^{-2k|z|} \Big(dt^{2} - dz^{2}\Big) - 
e^{2k(t - |z|)} (dy^{2} + dx^{2}),
\label{vismetric}
\end{equation} 

Here, $k = 2{\pi}{\sigma}$, and the wall is located at $z = 0$.
The gravitational field of this solution has amusing properties.  For
example, if you take the Newtonian limit of the Einstein equations for
(\ref{vismetric}) you obtain the equation
\[
{\nabla}^{2}{\phi} = -2{\pi}{\sigma}
\]
\noindent where $\phi$ is the Newtonian gravitational potential and $\sigma$
is the energy density of the wall.  From this equation it is clear that
a wall with {\it positive} surface energy density will have a repulsive
gravitational field, whereas a wall with negative energy density will
have an attractive gravitational field.  An even simpler way to see that
the (positive $\sigma$) VIS wall is repulsive is to notice that the
$t-z$ part of the metric is just the Rindler metric.

Further information is recovered by noticing that the $z$= constant
hypersurfaces are all {\it isometric} to $2+1$ dimensional de Sitter space:
\begin{equation} 
ds^{2} = dt^{2} - e^{2kt}(dy^{2} + dx^{2}). 
\end{equation} 

Given that $2+1$ de Sitter has the topology ${\rm S}^{2} \times {\Bbb R}$
it follows that the domain wall world sheet has this topology.  In other
words, at each instant of time the domain wall is topologically a 
two-dimensional sphere.  Indeed, in the original Ipser-Sikivie paper
a coordinate transformation was found which takes the $(t, x, y, z)$
coordinates to new coordinates $(T, X, Y, Z)$ such that in the new coordinates
the metric becomes (on each side of the domain wall):
\begin{equation} 
ds^{2} = dT^{2} - dX^{2} - dY^{2} - dZ^{2}.  
\end{equation}

\noindent Furthermore the domain wall, 
which in the old coordinates is a plane located at
$z = 0$, is in the new coordinates the hyperboloid
\begin{equation} 
X^{2} + Y^{2} + Z^{2} = {1\over {k}^2} + T^{2}.  
\label{dwhyper} 
\end{equation} 

Of course, the metric induced on a hyperboloid embedded in Minkowski spacetime
is just the de Sitter metric, and so this is consistent with what we have 
already noted.  This metric provides us with a useful way of constructing the
maximal extension of the domain wall spacetime:

\noindent First, take two copies of Minkowski space, and in each copy consider the
interior of the hyperboloid determined by equation
(\ref{dwhyper}), match these solid hyperboloids to
each other across their respective boundaries; there will
be a ridge of curvature (much like the edge of of a lens)
along the matching surface, where the domain wall is
located.
Thus, an inertial observer on one side of the wall will see the
domain wall as a sphere which accelerates 
towards the observer for $T<0$, 
stops at $T=0$ at a radius ${k}^{-1}$,
then accelerates away for $T>0$.  We illustrate this construction below,
where we include the acceleration horizons to emphasize the causal structure.
\vspace*{0.3cm}
\begin{figure}[htb]
\hspace*{\fill} \vbox{\epsfxsize=11cm
\rotate[r]{\epsfbox{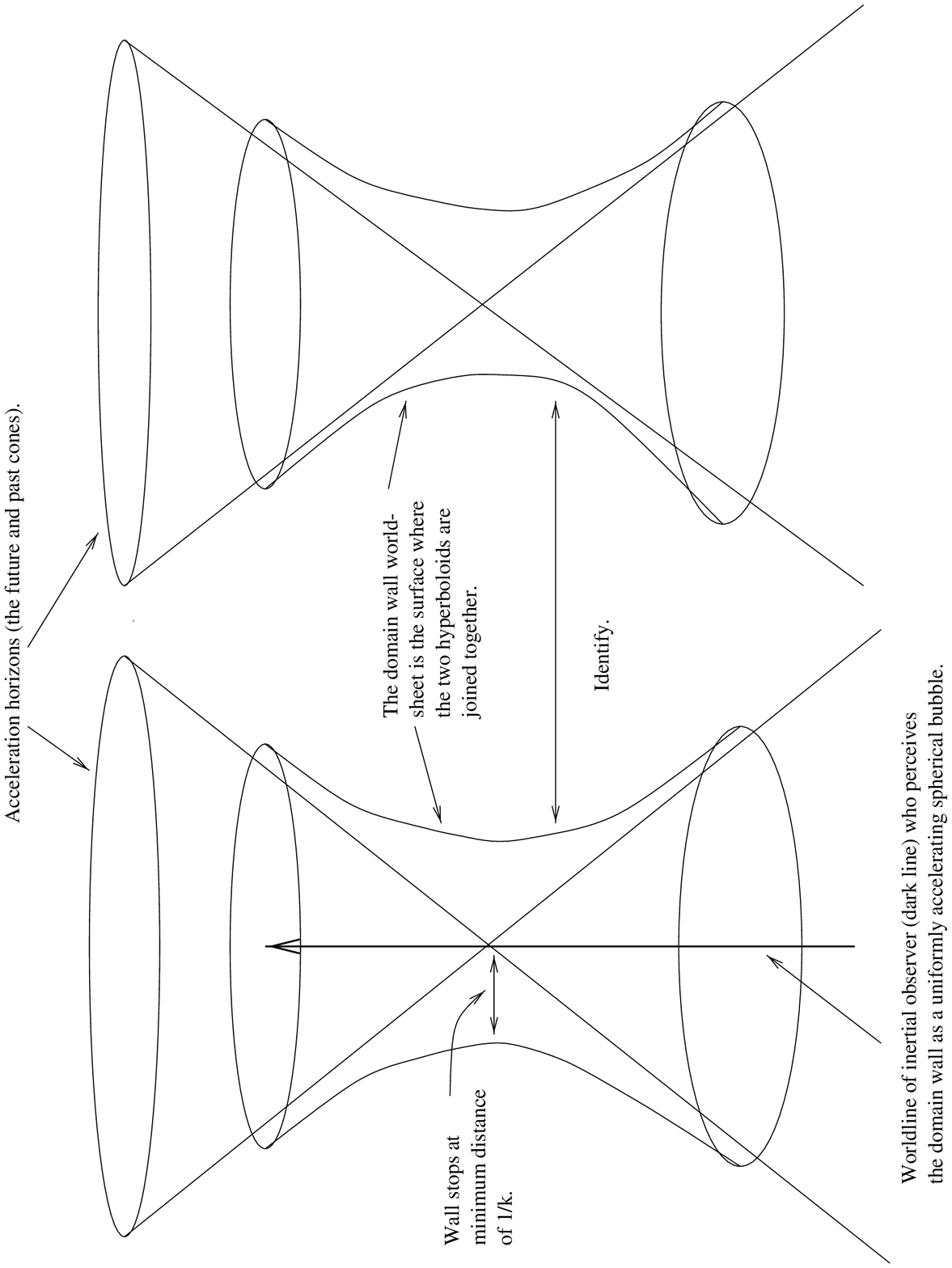}}} \hspace*{\fill}
\caption[Causal structure of VIS domain wall spacetime.]
{\small\sl}
\label{vis}
\end{figure}

Now, the repulsive effect of this vacuum domain wall is very similar to the
inflationary effect of a positive cosmological constant seen in de Sitter space.
Indeed, we often find it is useful to think of a VIS spacetime as an
inflating universe where all of the vacuum energy has been `concentrated'
on the sheet of the domain wall.

\section{`Popping' a VIS domain wall: Instanton and action}

\subsection{Euclidean section of the ingoing state}

Before we construct instantons for popping domain walls, it is useful if
we first recall a few basic facts about the Euclidean section 
and action of the ordinary VIS domain wall.

In the simplest scenarios, a VIS domain wall will form when 
there is a breaking of some discrete symmetry.  Usually,
one thinks of the symmetry breaking in terms of some Higgs field
$\Phi$.  If ${\cal M}_{0}$ denotes the `vacuum manifold' of $\Phi$
(i.e., the submanifold of the Higgs field configuration space on which
the Higgs acquires a vacuum expectation value because it will minimize
the potential energy $V(\Phi)$), then a necessary condition for a
domain wall to exist is that ${\pi}_{0}({\cal M}_{0}) \not= 0$.  In
other words, vacuum domain walls arise whenever the vacuum manifold is
not connected.  Given these assumptions, one usually writes 
the Lagrangian density for the matter field $\Phi$ as
\begin{equation}
{\cal L}_{m} = 
-{\frac{1}{2}}g^{{\alpha}{\beta}}{\partial}_{\alpha}
{\Phi}{\partial}_{\beta}{\Phi} -
V(\Phi).
\end{equation}

The exact form of $V(\Phi)$ is not important.  All that we
require in order for domain walls to be present is that $V(\Phi)$ has
a discrete set of degenerate minima, where the potential vanishes.
Given this matter content, the full (Lorentzian) Einstein-matter
action then reads:
\begin{equation}
S = {\int}_{\!\!M} d^4\! x \, \sqrt{-g}\,
\Big[ {R \over 16 \pi} + {\cal L}_{m} \Big] + 
\frac{1}{8{\pi}}{\int}_{\!\!\partial M} d^{3}\!x\, \sqrt{h}K.
\end{equation}

Here, $M$ denotes the four-volume of the system, and $\partial M$
denotes the boundary of this region.  One obtains the Euclidean
action, $I$, for the Euclidean section of this configuration by
analytically continuing the metric and fields and reversing the
overall sign.  The `simplified' form of this Euclidean action in the
thin wall limit has been derived in a number of recent papers
(\cite{cald}, \cite{shawn1}) and so we will not reproduce the full
argument here.  Basically, one first {\it assumes} that the
cosmological constant vanishes ($R = 0$) and then one uses the fact that the
fields appearing in the matter field Lagrangian depend only on the
coordinate `$z$' normal to the wall, and one integrates out this
$z$-dependence to obtain the expression
\begin{equation}
I = - \sum_{i=1}^{n}
 {\frac{\sigma_{i}}{2}}{\int}_{\!\!D_{i}}d^{3}\!x \sqrt{h_{i}}.
\end{equation}

Here, $D_i$ denotes the $i$-th domain wall, ${\sigma}_i$ is the energy
density of the domain wall $D_i$, $h_i$ is the determinant of the
three-dimensional metric ${h^{ab}}_{(i)}$ induced on the domain wall
$D_i$ and $n$ is the total number of domain walls.  It is not hard to 
prove that variation relative to ${h^{ab}}_{(i)}$ on each domain wall
will yield the Israel matching conditions.

Now, as we have seen the Lorentzian section of a single VIS domain wall
is just two portions of flat Minkowski space glued together.  It is 
therefore natural to propose that the Euclidean section is obtained
by gluing two flat Euclidean four-balls together along a common
$S^3$ boundary component.  In this way one obtains the `lens instanton',
which describes (in the context of the no-boundary proposal) the creation
of a single VIS domain wall from `nothing'.  This lens instanton is
the Euclidean section of the incoming state - the VIS domain wall
before the decay process (puncture) has taken place.  In order to calculate
the rate for the process, we now need to construct the instanton for the
perforated domain wall.

\subsection{Euclidean section of the outgoing state}

As we saw above, a VIS domain wall moving in imaginary time sweeps out
an $S^3$ `ridge' of curvature along the equator of the lens instanton.
In a similar way, we expect a virtual loop of string moving
in imaginary time to sweep out a two-sphere, $S^2$.  For our purposes,
we want this string loop to correspond to a puncture which appears
on the Euclidean portion of the domain wall, expands to some maximum
size, then collapses again.  In other words, the Euclidean section of
the punctured domain wall is obtained by taking the lens instanton,
and `truncating' each flat four-ball so that the `hole' swept out by 
the string is made manifest.  This is illustrated below:
\vspace*{0.3cm}
\begin{figure}[htb]
\hspace*{\fill} \vbox{\epsfxsize=11cm
\rotate[r]{\epsfbox{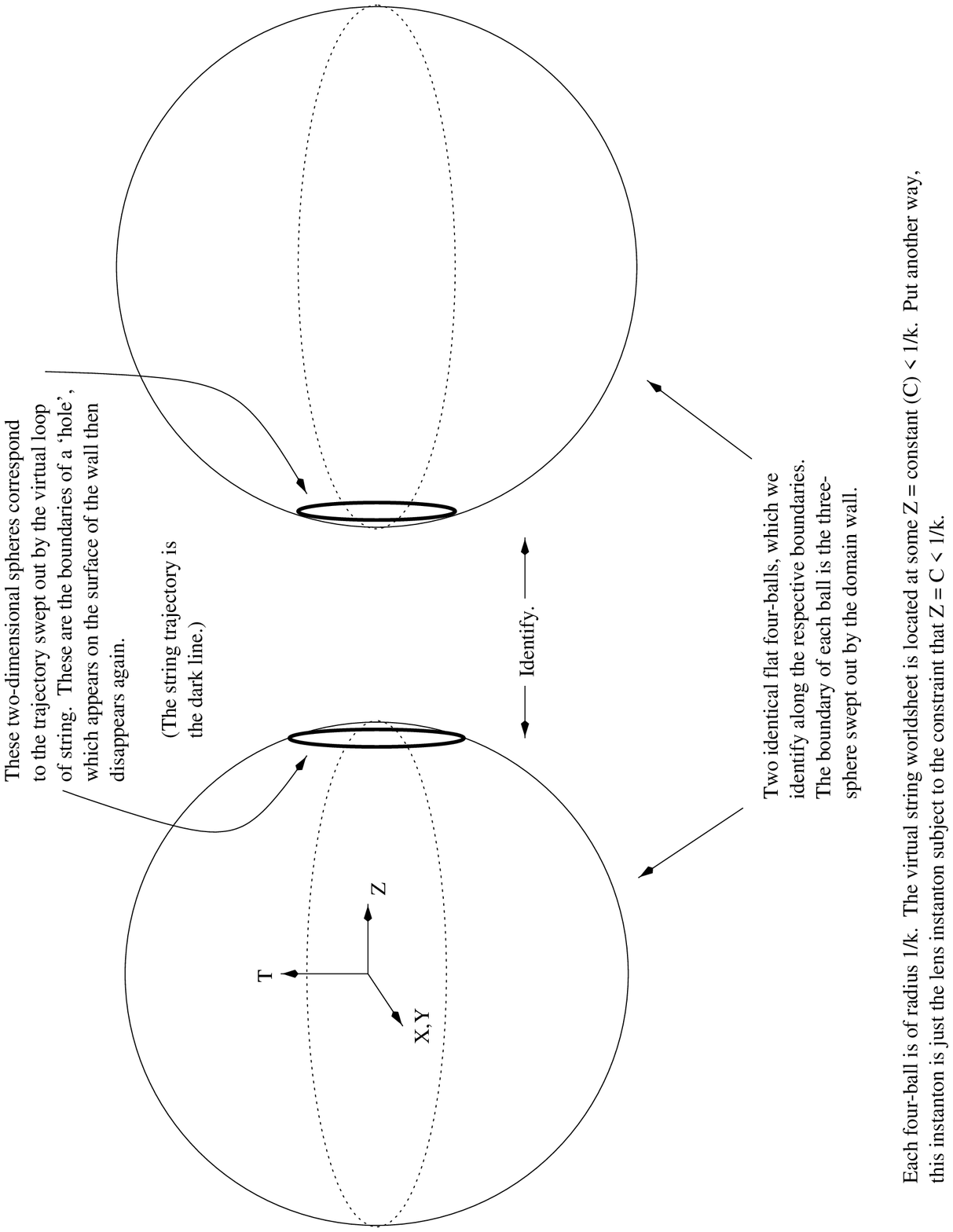}}} \hspace*{\fill}
\caption[Euclidean section for a punctured domain wall.]
{\small\sl}
\label{puncture}
\end{figure}

More explicitly, if we write the metric on flat space in the usual form
\[
ds^2 = d{\cal T}^2 + dX^2 + dY^2 + dZ^2
\]

\noindent where ${\cal T} = iT$ denotes imaginary time, then the domain wall
(on the lens instanton) is located at the boundary of each ball of radius
$R = 1/k$.  In order to obtain the instanton for the punctured wall, we
truncate the range of the variable $Z$:
\[
Z ~{\leq}~ C < R
\]

\noindent where $C$ is some constant.  In other words, we
shave off a portion of the lens instanton so that the virtual string
worldsheet is located at the surface $Z = C$.

\subsection{Calculation of the action and amplitude}

We are now in position to calculate the action and decay rate for this
process.  Since there is no topology change in this situation, it is
easy to construct an interpolating instanton by finding 
slices of the initial instanton which match smoothly to
slices of the final instanton.  We may therefore use the no-boundary
ansatz without relying on more sophicticated arguments, such as
the `patching proposal' put forward in \cite{patch}.  As described
in the introduction, we want to calculate the action difference,
${\Delta}S$, between the Euclidean action of the final configuration
($S_F$) and the Euclidean action of the initial configuration ($S_I$).

Since the final instanton is obtained from the initial instanton simply
by removing a `cap' from the boundary of each four-ball before performing
identifications, it follows that the only action difference can come
from this contribution.  As in the non-gravitating case described in
the introduction, the removed region will contribute a bulk term
(corresponding to the hole in the domain wall) and a surface term
(the boundary of the hole).

The volume of the removed cap is calculated to be
\[
2{\pi}^{2}R^{3}\Big(1 - \frac{2}{\pi}(\frac{-C}{2R}{\sqrt {1 - C^{2}/R^2}}
+ \frac{cos^{-1}(C/R)}{2}) \Big)
\]

\noindent Likewise, the surface swept out by the string is just a 
two-sphere of radius $(R^2 - C^2)^{1/2}$.  Weighting the bulk term
with domain wall energy density $\sigma$, and the surface term
with string tension $\mu$ we therefore obtain
\begin{equation}
{\Delta}S = S_F - S_I = 4{\pi}(R^2 - C^2){\mu} - 
{\pi}^{2}R^{3}\Big(1 - \frac{2}{\pi}(\frac{-C}{2R}{\sqrt {1 - C^{2}/R^2}}
+ \frac{cos^{-1}(C/R)}{2}) \Big) {\sigma}
\end{equation}

\noindent Ignoring the prefactor term, the decay probability
is then given as
\begin{equation}
{\cal P} = e^{-{\Delta}S}
\end{equation}

\noindent (Actually, the prefactor term may contain interesting information
in this situation, for the simple reason that when the domain wall
`pops', the overall symmetry of the spacetime is broken from spherical
symmetry to cylindrical symmetry; presumably, fluctuations about the
instanton would respect this symmetry breaking.  We will have more to
say about the evolution of quantum fields in a punctured domain wall
spacetime later in this paper).

Now, just as for popping light domain walls \cite{review}, our use of
the thin wall approximation to derive (3.4) is only justified if the scale
of symmetry breaking for the strings is {\it much} larger than the 
scale of symmetry breaking for domain walls.  It follows that, generically,
we will have ${\Delta}S >> 1$, and hence ${\cal P} << 1$.

\subsection{Decay of domain walls with multiple punctures}

It is also possible for several holes to spontaneously form on the 
surface of a domain wall.  In the simplest situation, all of the 
string loop boundary components will nucleate at the same initial
radius ($(R^2 - C^2)^{1/2}$), with the same string tension $\mu$
so that the total action will just be
\[
{\Delta}S_{TOT} = N {\Delta}S
\]

\noindent where ${\Delta}S$ is given by (3.4), and $N$ is the total
number of holes.  Of course, one might also imagine more exotic scenarios
where the holes nucleate at different initial radii $R_{i} = (R^2 - {C_{i}}^2)^{1/2}$,
and string tension ${\mu}_i$, so that the total action for the decay would
be given as
\[
{\Delta}S_{TOT} = \sum_{i=1}^{N} {\Delta}S_{i}
\]

\noindent where
\[
{\Delta}S_{i} = 4{\pi}({R}^2 - {C_i}^2){\mu}_{i} - 
{\pi}^{2}R^{3}\Big(1 - \frac{2}{\pi}(\frac{-C_i}{2R}{\sqrt {1 - {C_i}^{2}/R^2}}
+ \frac{cos^{-1}({C_i}/R)}{2}) \Big) {\sigma}
\]

\noindent In either situation, the creation of multiple holes is heavily
suppressed relative to the nucleation of a single hole.

Of course, it is of some interest to know whether or not a domain wall
can ever be completely annihilated by the processes which we are discussing here.
In fact, it is not hard to prove that a domain wall will be completely
destroyed by this decay process whenever at least {\it four} string loop
boundary components are nucleated.

In order to understand this, recall that the worldvolume of the domain wall
is isometric to $2+1$-dimensional de Sitter spacetime (embedded in 
$3+1$-dimensional Minkowski), and that the trajectory of a given string
loop boundary component may be obtained by taking the intersection of
the hyperboloid with a surface of constant $Z$ (relative to the coordinates
(2.4) on Minkowski space), which is simply a copy of $2+1$ dimensional
Minkowski spacetime.  If we view all of this from `above' the hyperboloid,
then we see that the domain wall is a two-sphere which expands uniformly
outwards, and that the string loop boundary components are intersections
of this two-sphere with flat timelike hypersurfaces.  If there are at least four
(non-parallel) such hypersurfaces, then on each spacelike surface the
hypersurfaces will bound a tetradedron of fixed size.  In three dimensions,
a tetrahedron may bound a two-sphere, and so initially at least the domain
wall may still lie (partially) within the tetrahedron.  However, if the 
sphere continues to expand it will always eventually envelop the tetrahedron,
and hence the domain wall will have been annihilated (i.e., the string
boundary components `collide' precisely at the corners of the tetrahedron).
If you only had three (or fewer) timelike hypersurfaces, you could never
completely bound the two-sphere in this way.  The intersection of
the sphere with a given timelike hypersurface would continue unbounded
in some direction (because you would never encounter another timelike
hypersurface).  Thus, at least a portion of the domain wall would
survive eternally.  In other words, you need to nucleate at least four punture wounds
in a domain wall in order to completely annihilate the domain wall.

\section{Lorentzian evolution of punctured domain walls}

In the last section, we showed that a VIS domain wall may decay via
the formation of closed string loop boundary components on the worldvolume
of the wall.  We found instantons, and calculated the corresponding
actions and rates, for such processes.  We now turn our attention
to the Lorentzian, or `real time', picture of this process.

First, recall the representation (Fig. 1) of the VIS domain wall
as two solid hyperboloids in Minkowski space, identified along their
respective boundaries.  The constraint on the coordinated $Z$,
$Z ~{\leq}~ C < R$, which we imposed on the Euclidean section, extends
to the Lorentzian section as well.  Thus, in the Lorentzian coordinates
$(T,X,Y,Z)$, the equation of motion for the loop of string may be written as
\begin{equation}
X^2 + Y^2 = (R^2 - C^2) + T^2
\end{equation}

\noindent Thus, the initial radius of the loop of string (at $T = 0$)
is seen to be $R^2 - C^2 > 0$, which is what we expect given that the Lorentzian
section must match smoothly to the Euclidean section.  At late times,
the hole is expanding at the speed of light.  Of course, the hole never
completely devours the wall for the simple reason that the spherical
wall also expands exponentially.  The global structure of this 
spacetime is illustrated below:
\vspace*{0.3cm}
\begin{figure}[htb]
\hspace*{\fill} \vbox{\epsfxsize=11cm
\rotate[r]{\epsfbox{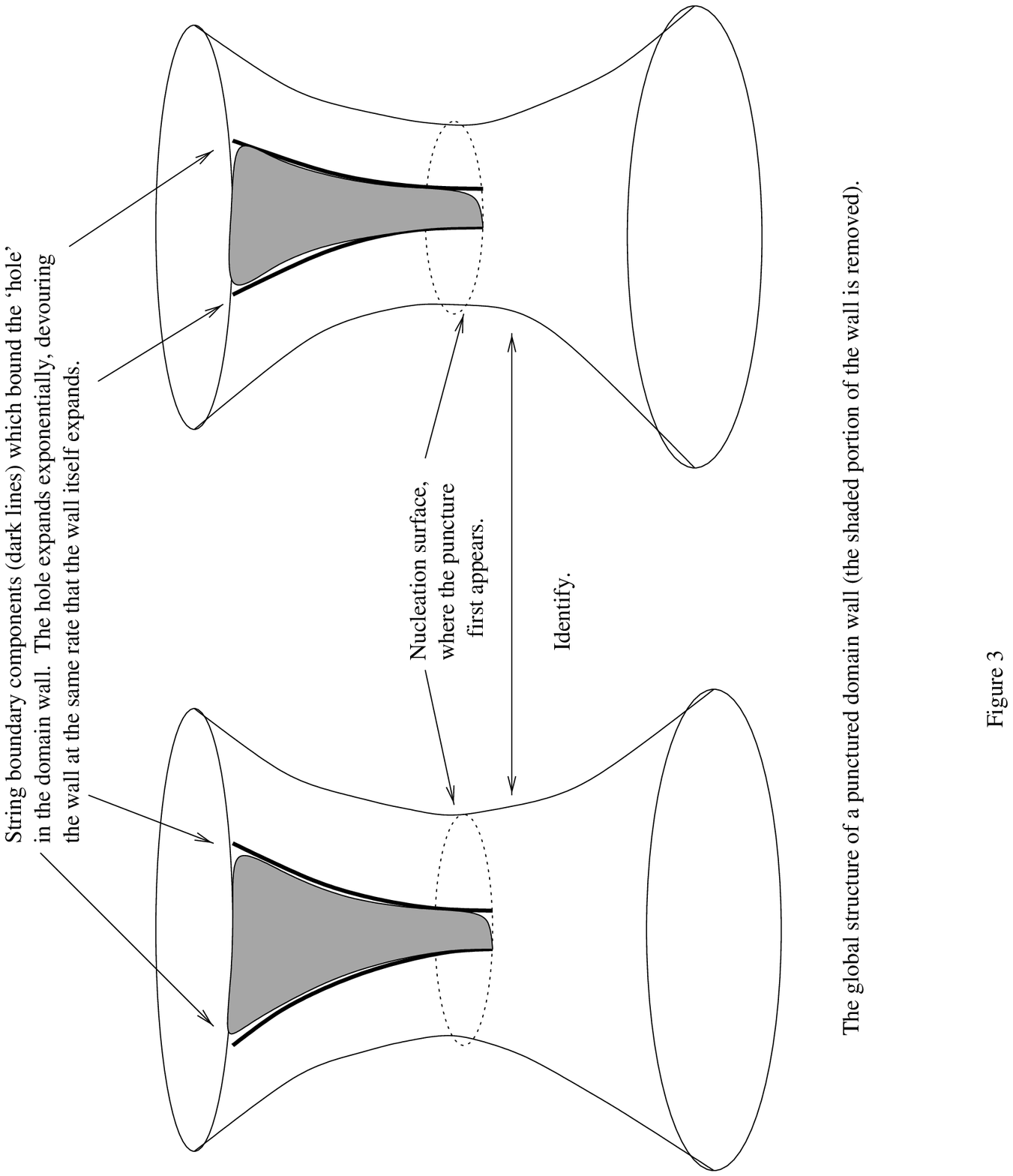}}} \hspace*{\fill}
\caption[Global structure of the punctured VIS domain wall spacetime.]
{\small\sl}
\label{pop}
\end{figure}

It is interesting to consider the gross properties of particles which
are propagating in the background of a wall which spontaneously decays
in this way.  For example, suppose we consider some scalar field $\psi$
which couples to the Higgs field (of the domain wall) $\phi$ through the
interaction
\[
{\cal L}_{int} = -\frac{\lambda}{2}{\phi}^{2}{\psi}^{2}
\]

\noindent (Here, we are thinking of $\phi$ as a Higgs field with a standard
${\phi}^4$, `double-well' potential).  In \cite{review} it is shown
that the reflection coefficient for scattering of ${\psi}$ particles
off of a $\phi$ domain wall is {\it zero} in the limit that the domain
wall has zero thickness.  In other words, it makes sense to think of the
(infinitely thin) VIS domain walls which we have been considering in
this paper as spherical, accelerating, `moving mirrors'.  Put another
way, each side of the domain wall is a spherical cavity, with Dirichlet
boundary conditions for the fields at the boundary of the cavity.

In \cite{bir} it is shown that a mirror moving with uniform acceleration
$a$ will generate (relative to an inertial particle detector) a thermal
bath of radiation at temperature
\[
T = \frac{a}{2\pi k_{b}}
\]

\noindent where $k_b$ is Boltzmann's constant.
Of course, we could have predicted that the VIS spacetime would
have an entropy and temperature of this form, simply because the repulsive
energy of the wall generates a cosmological horizon which leads to loss
of information.  Inertial observers in a VIS spacetime can never recover information
about what is happening on the other side of the domain wall, and they
will represent this ignorance by tracing over states associated with the
horizon.

If a domain wall decays by the formation of a closed string
loop boundary component on the worldvolume of the wall, the isotropic
thermal nature of the initial (VIS) spacetime should be lost.  This is because
the initial spherical symmetry will be broken to cylindrical symmetry
when the hole forms.  The hole is in some sense then a `window' between
the two spherical cavities, through which information can propagate.
It would be interesting to have some explicit calculation for the 
scattering of $\psi$-particles in the background of a VIS domain wall
with a single puncture.

As we have shown, if four (or more) punctures spontaneously nucleate on
a wall, the string boundary components must eventually collide and
annihilate the wall, so that all of the energy density initially stored
in the wall is thermalized.  The final `annihilation' of a domain wall
by this process is rather analagous to the endpoint of inflation,
in the sense that the source driving the exponential expansion is suddenly
`switched off'.  However, a domain wall with several punctures will
presumably generate a highly non-isotropic radiation background, in contrast
with the inflationary scenario.

\section{Conclusions: A potential brane-world instability?}

We have shown how to describe the decay of vacuum domain walls by the
formation of closed string boundary components, when the effects of gravity
are included.  We found new instantons, as well as the corresponding Lorentzian
solutions, which describe the formation and evolution of multiple closed
string loop boundary components on a given VIS domain wall background.
It would be nice to have a complete picture of how quantum fields will
evolve in the background of one of these punctured walls.

In general domain walls arise as (D-2)-dimensional defects
(or extended objects) in D-dimensional spacetimes.  In fact, domain
walls are a common feature in the menagerie of objects which appear
in the low-energy limit of string theory, as has been discussed in
detail in \cite{paul} and \cite{cvet}.  

Of course, there has recently been an enormous amount of interest
in the possibility that the universe itself might be a domain wall
moving in some five-dimensional bulk spacetime.  In particular, Randall
and Sundrum \cite{rs} have recently put forward a model where the universe
is a ${\Bbb Z}_2$-symmetric positive tension domain wall bounding two 
bulk regions of anti-de Sitter (adS) spacetime.  In their original scenario,
the bulk cosmological constant is fine-tuned relative to the domain wall
energy density so that the effective cosmological constant on the brane
is precisely zero.  However, during a period of inflation on the brane-world
the effective cosmological constant would be positive, and the domain wall
would simply be a four-dimensional de Sitter hyperboloid embedded in
five-dimensional adS (for a recent interesting discussion of various
semiclassical instabilities associated with these de Sitter brane-worlds
see \cite{garriga}).  Thus, the causal structure of these de Sitter
brane-worlds is identical to that of the VIS domain wall \footnote{The
causal structure of these walls, in the four-dimensional case, 
was first discussed in \cite{mirjam}; it was shown in \cite{mirjam2}
that this causal structure is universal in any dimension}.  

Now, one can certainly {\it imagine} that the domain wall of the Randall-Sundrum
model arises because of a symmetry breaking pattern which allows for the
universe itself to end on a two-brane (all that is required is an associated
exact sequence of the form (1.1)).  In such a scenario, the brane-world would
be unstable to the formation of `holes' - two-dimensional surfaces where the
universe would end.  These holes would accelerate out, devouring the universe
and converting brane-world fields into bulk degrees of freedom.  Since a
fundamental polytope in four dimensions has five faces, it follows that you
would have to nucleate at least five of these holes to completely annihilate
a de Sitter universe in this fashion.  

One could even go further and
imagine scenarios where the brane-world is {\it itself} the boundary of a
puncture wound in some four-brane (in this way you could have more than
one large extra dimension).

In general, we would expect
many of the (self-gravitating) p-branes of supergravity models
to be unstable to the sort of decay processes which
we have described in this paper.  From this point of view, the process which
we have studied is just another example of the sort of
`brane damage' which we expect to be a generic feature of 
the p-branes of M-theory.

Research on these and related questions is currently underway.

{\noindent \bf Acknowledgements}\\

The authors thank  Richard Battye, Robert Caldwell, Sean Carroll, 
Gary Gibbons and James Grant for useful conversations.  A.C. was supported
by a Drapers Research Fellowship at Pembroke College, Cambridge,
and is currently (partially) supported at MIT
by funds provided by the U.S. Department of Energy (D.O.E.) under
cooperative research agreement DE-FC02-94ER40818.  A.C. also thanks
the organizers and participants of the ITP Program on Supersymmetric Gauge
Dynamics and String Theory, where this work was completed, for stimulating
discussions.  At ITP A.C. was supported by PHY94-07194.

\end{document}